# Martini coarse-grained model for clay-polymer nanocomposites

*Parvez khan, Gaurav Goel**

Department of Chemical Engineering, Indian Institute of Technology Delhi, New Delhi-110016, India

ABSTRACT

We have developed a coarse-grained (CG) model of a polymer-clay system consisting of organically modified montmorillonite nanoclay as the nanoparticle in accordance with the MARTINI forcefield. We have used mechanical properties and cleavage free energy of clay particle to respectively parameterize bonded and nonbonded interaction parameters for an organically modified montmorillonite (oMMT) clay particle where intergallery $Na^+$ ions are replaced by tetramethylammonium (TMA) ions. The mechanical properties were determined from the slope of stress-strain curve and cleavage free energy was determined by allowing for full surface reconstruction corresponding to a slow equilibrium cleavage process. Individual dispersive and polar contributions to oMMT cleavage energy were used for determination of appropriate MARTINI bead types for CG oMMT sheet. The self-consistency of developed MARTINIFF parameters for TMA-MMT – polymer system was verified by comparing estimates for select structural, thermodynamic, and dynamic properties obtained in all-atomistic simulations with that obtained in coarse-grained simulations. We have determined the influence of clay particle on properties of three polymer melts (polyethylene, polypropylene, and polystyrene) at two temperatures to establish transferability of the developed parameters. We have also shown that the effect of clay-polymer interactions on structure-property relationships in polymer-clay nanocomposite system is well captured by Rosenfeld's excess entropy scaling.

**1. Introduction**

Organic-inorganic hybrids are a versatile class of materials whose mechanical, thermal, and barrier properties can be tuned for a diverse set of applications in areas ranging from space operation, consumer goods to health, medical and food technology.[1,2] Polymer nanocomposites (PNCs), one such important hybrid material, often exhibit excellent mechanical, thermal, electrical, and optical properties, which in-turn are intricately linked to the overall morphology of the composite.[3,4] Therefore, polymer materials reinforced with nanoparticles, ranging from spherical metallic or latex nanoparticles, nanotubes such as CNTs, sheet/platelet nanoparticles such as grapheme sheet, phyllosilicates, have received significant attention in both fundamental and applied sciences.[5,6] Here, we have investigated PNCs consisting of highly anisotropic layered-silicate (clay) nanoparticles.[7] The properties of such polymer-clay nanocomposite (PCNC) are intricately linked to the extent to which clay tactoids exfoliate and disperse in the polymer matrix during processing.[8–11] A large number of material and process parameters, and non-monotonic dependence of target properties (e.g., gas and vapor permeability for packaging applications) on these parameters makes the development of PCNCs for specific applications a challenging task.

Both molecular dynamics simulations and experimental characterization have shown that composite morphology and properties depend rather sensitively on chemical details of the system,[12,13] with solubility parameters for various PNC components, and clay modifier length and density on platelet surface identified as important parameters governing composite morphology and properties.[14–16,17–21] Further, it has been suggested that the length scale associated with composite properties, e.g., water vapor permeability properties of polymer films, is not the mesoscale but rather the nanoscale of the material.[22] Whereas mesoscale and continuum simulation (finite element method, dissipative particle dynamics) methods find use in estimation of macroscopic properties consistent with modelled effective potentials, molecular simulations with chemically specific forcefields are better suited to investigate consequences of specific intermolecular interactions between various PNC components. Further, equilibrium molecular simulation trajectories are required for accurate determination of these effective potentials. However, the key issue that limits the application of molecular simulation on polymeric materials is the long relaxation times associated with polymer dynamics and the requirement of large system size for anisotropic particles. To this end, systematic coarse-grained (CG) models have been developed for various macromolecules, wherein a few atoms are replaced by a single bead and consistent effective interaction potentials are obtained for such CG beads. One category of CG forcefields have modeled polymer chain by a bead-spring model with non-bonded interactions modeled at a low-level detail as "repulsive", "neutral", or "attractive".[23,24] In another class of CG forcefields more details of the specific chemical make-up of the effective bead are included enabling investigation of effect of nature of monomer-monomer, monomer-solvent and/or monomer-particle interactions.[25–30] These effective potentials between coarse-grained beads can be obtained by either employing an iterative Boltzmann inversion (IBI) scheme to match radial distribution functions[31] or by determining potentials of mean force between CG beads.[32] Johnston and Harmandaris[11] have reviewed work on development of CG potentials and multiscale modeling approaches for model polymer-solid interfacial systems.

Chemically specific coarse-grained models for the specific case of silicate mineral clay – polymer composites (PCNCs) [9,33–35] can be extremely useful because of significant effect of the nature of interactions between various components on composite properties.[36] It has been shown that a subtle management of characteristics (polarity, aspect ratio, molecular weight, etc.) of individual components can lead to a synergistic improvement in properties of PCNCs.[14,16] Suter et al. obtained a coarse-grained yet chemically specific representation of interaction potentials for a montmorillonite-poly(ethylene glycol)/poly(vinyl alcohol) composite by employing the IBI scheme and were able to provide a quantitative description of dynamical process of polymer intercalation into clay tactoids.[33] Structure-based IBI schemes have been shown to often have poor temperature transferability of interaction parameters[37] and are expected to require additional parameterization on change of polymer matrix or clay intergallery ions. [33,38]

In this work we use an alternative approach wherein we have extended the existing MARTINI forcefield (MARTINIFF) to polymer-clay systems. MARTINIFF is a thermodynamics-based CG model wherein 3-4 atoms are grouped into a single bead, and the interaction parameters are determined by reproducing the densities and partitioning free energy of molecules between an aqueous phase and an organic phase.[39,40] The MARTINIFF beads are classified into four main types on the basis of their polarity. It was originally developed for lipids, and has been shown to be accurate and up to 200 times faster than all-atomistic simulation.[40] It was shown to capture 75-100% of major collective motions for lipid chains with 5-7 fold faster sampling, albeit it overestimated short-range ordering.[41] MARTINIFF has been parameterized for several homogenous polymer systems, e.g. polyethylene glycol (PEO)[42], polystyrene (PS)[29], polyethylene (PE) and polypropylene (PP)[28]. More recently, it was used for heterogeneous systems to investigate different phenomenon such as mechanism of adsorption of detergents onto a carbon nanotube[43], preferential adsorption of long-chain organic molecules on graphite surface from solvent and the formation of ordered structures on the surface through self-assembly[44] and the structural changes to lipid membrane that result from the addition of aliphatic alcohols with various alkyl tail lengths.[45] MARTINIFF is based on parameterization of individual chemical building blocks, and therefore, provides a systematic framework for extension to multi-component systems. Transferability of MARTINIFF parameters with change in temperature, solvent, or particle functionalization has been validated for a variety of systems.[29] We have used mechanical properties and cleavage energy of clay particle to respectively parameterize bonded and nonbonded interaction parameters for an organically modified montmorillonite (oMMT) particle. Individual dispersive and polar contributions to oMMT surface energy were used for determination of appropriate MARTINI bead types for CG oMMT sheet. The surface energy is determined by allowing for full surface reconstruction corresponding to a slow equilibrium cleavage process.[46] A self-consistent parameter set for clay-polymer interactions is then obtained to accurately reproduce structural, thermodynamic, and dynamic properties of a polymer-clay system. The transferability of developed parameter set was verified by varying the matrix polymer (PP, PE, or PS) and temperature. Structure-property relationships in polymer-clay systems were investigated using Rosenfeld scaling.

## 2. Methods

Fig. 1(a) shows the 2:1 (T-O-T) structure of montmorillonite (MMT), typical of layered phyllosilicates. We have used organically modified MMT clay particle with unit cell formula [TMA]$_{0.33}$[Si$_4$O$_8$][Al$_{1.667}$Mg$_{0.333}$O$_2$(OH)$_2$], where MMT intergallery ions have been replaced by tetra-methyl ammonium (TMA) ions. Isomorphic substitution of Al$^{3+}$ by Mg$^{2+}$ in the octahedral layer were chosen at random to obtain a charge exchange capacity (CEC) of 90 mmol/100 g and it was ensured that no Mg–(OH)$_2$–Mg pairs exist.[47,46] Three different polymers were used, namely polyethylene (molecular weight: 2240 g/mol; PE80), polypropylene (4200 g/mol; PP100), and polystyrene (10400 g/mol; PS60). The molecular weight was chosen to be just below the entanglement length of each polymer[48] so that equilibrium trajectories are sampled in reasonable simulation time.

Both all-atomistic (AA) and coarse-grained (CG) molecular dynamics simulations were carried out using GROMACS 5.1.4.[49,50] Packmol[51] was used for creating initial box configuration, and Chemdraw software was used for generating initial conformation of polymer chains. For AA simulations, we have used CHARMM36 FF[52,53] for the polymer and TMA ions, and the INTERFACE FF[46,54] for MMT. All the cross-interactions were defined by standard mixing rules.[46] The INTERFACE FF has been shown to provide very good agreement for a variety of properties when compared to experimental data. Some key results include excellent reproduction of basal spacing of Na-MMT and C$_{18}$-MMT, cleavage energies of Na-MMT[46]. The MARTINI[55] FF was used for coarse-grained simulations. MARTINIFF parameters for PE, PP, and PS were taken from previous work on parameterization of polymer melts.[28,29] MARTINIFF parameters for the MMT beads as well as MMT-polymer cross-interaction have been developed in this study. We have used a successive set of simulations with increasing details such that maximum two parameters are fitted in a given simulation. The self-consistency and transferability of developed parameters was validated by comparison of structural, thermodynamic, and dynamic properties in AA and CG simulations at two different temperatures.

A time step of 2 fs was used for AA simulations and 10 fs for CG simulations. For AA simulations, a cutoff of 1.2 nm with force-shift of 1.0 nm was used for VDW interactions and PME $\left(\text{accuracy } 10^{-2} \frac{\text{kcal}}{\text{mol}}\right)$ was used for long-range Coulombic interactions. For CG simulations, a cutoff of $r_{\text{cut}} = 1.1$ nm was used for non-bonded interactions represented by the 12-6 Lennard-Jones potential.[56] For the present study, high charge on the clay particle and low dielectric medium (polymer) necessitated use of PME for long-range Coulombic interactions. A relative dielectric constant of $\varepsilon_{rel} = 2.5$ was used to mimic low dielectric medium. For all simulations, we have used V-rescale for coupling the simulation box to a thermostat ($\tau_T = 0.1$ ps). The simulation box was coupled to a Berendsen[57] barostat during equilibration ($\tau_P = 2.5$ ps) and Parrinello-Rahman[58] barostat ($\tau_P = 3.0$ ps) for simulations used for property calculation. For in-vacuo (no solvent) simulations of clay particles, the box dimension along z-axis was fixed at 100 nm (to eliminate effect of electrostatic interactions with mirror image) with periodic boundary conditions along $x$ and $y$.

**2.1. Model development for coarse-grained TMA-MMT.** The chemical structure of MMT is such that $Si$, $Al$, or $Mg$ atoms are connected through M-O bonds ($M = Si, Al, or\ Mg$). Accordingly, we have defined each CG MMT bead as consisting of three heavy atoms, viz. $SiO_2$ or $MO_2H$ ($M = Al\ or\ Mg$), and CG bead center was put at $Si, Al, or\ Mg$ atom. This representation allowed us to preserve the 2:1 layered structure of phyllosilicates through interlayer and intralyer bonds between CG beads (Figure 1). We have used the smaller (fine grained) type-S representation for all three beads, as found to be suitable for ring structures in MARTINI parameterization.[29,40,44]

**Bonded Potential.** We have used a single sheet of TMA-MMT placed in $xy$ plane with bond connections across the periodic boundary (periodic sheet). The equilibrated structure of Na-MMT of size 3.11 nm × 2.71 nm (924 atoms in 6 × 6 × 1 array of unit cells) was taken from INTERFACE FF database and intergallery Na$^+$ ions were replaced by TMA ions. The position of TMA ions was then equilibrated by moving them 1 nm away from both (100) surfaces, creating several different (10) initial arrangements and carrying MD simulations till potential energy converges for each starting structure. The configuration with lowest potential energy was used for subsequent simulations. This equilibrated AA sheet was also used to obtain an initial configuration for the CG TMA-MMT sheet with 228 beads.

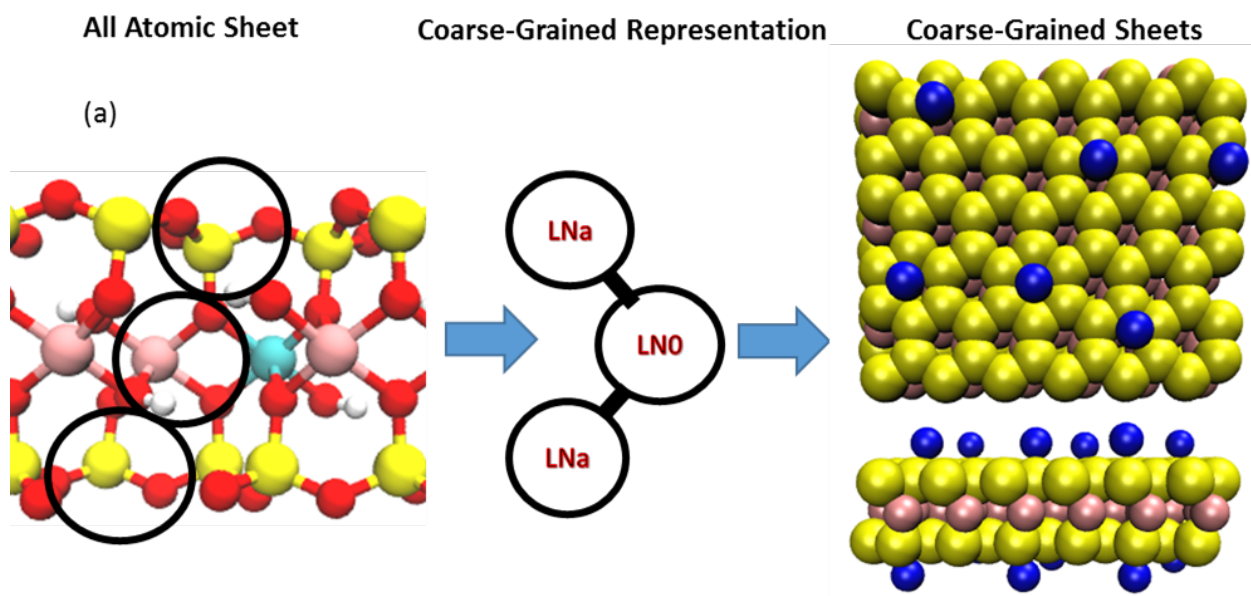

**Figure 1.** A schematic representation of multiscale model for Montmorillonite sheet. $SiO_2$ groups of tetrahedral layers are represented as LNa (corresponding to Martini SNa bead) and or $MO_2H$ ($M = Al\ or\ Mg$) groups of octahedral layer are represented as LN0 or LQ0 (corresponding to Martini SN0 and SQ0 beads). Self-interaction and cross interaction between LN0, LNa and LQ0 developed in this work. Cross interactions between these beads and Martini beads kept same as original Martini interactions.

The equilibrium statistics on bond and angle distribution for effective CG beads were obtained from a 10ns simulation of an all-atomistic TMA-MMT sheet. We have defined a bond potential for all neighboring CG bead pairs where the corresponding central atom ($Si, Al, or\ Mg$) are connected by an angle potential in AA representation. We have not used any torsion potential for the coarse-grained sheet since even the all-atomistic INTERFACE forcefield does not require any torsion potential to represent the clay particle.[46] We have used harmonic bond and angle potentials for CG beads with stiffness determined such that chosen mechanical properties obtained from AA simulations are reproduced by CG parameters.

***Determination of mechanical properties***. Young's modulus ($E_x$) and Poisson ratio ($\nu_{12}$) for the sheet were calculated from a series of simulations in which strain was increased by stepwise uniaxial extension (+) or compression (−) along $x$, according to $x_n = (1 \pm n\varepsilon_x)x_0$ where $x_n$ is box length along $x$ in the $n^{th}$ simulation, $x_0$ is equilibrium length, and $\varepsilon_x = 0.001$ (see Figure S1). The box dimension was increased up to 3.203 nm in extension runs and decreased up to 3.016 nm in compression runs. For this set of simulations, pressure coupling was used only along $y$. Relaxation along $y$ and a vacuum space above clay particle (along $z$) ensures null contribution of orthogonal directions to the stress-strain behavior. A 450 ps simulation was carried out at each box length with average stress and clay (or box) $y$-dimension calculated from the last 300 ps. $E_x$ was calculated from the $\sigma_{xx}$ component of stress tensor. As per Voigt notation, $\nu_{yx} = -\frac{\epsilon_y}{\epsilon_x}$, where $\epsilon_y$ is the resulting deformation in transverse direction because of applied deformation ($\epsilon_x$). Analogously, $E_y$ and $\nu_{xy}$ were calculated from extension or compression along $y$. Finally, thin plate relationship[59] is used for determination of bending stiffness ($D_x$) as per equation 1.

$$\frac{D_x}{h^2} = \frac{E_x h}{12(1 - \nu_{xy}\nu_{yx})} \tag{1}$$

where, $h$ is the thickness of the clay sheet defined as the distance between center of mass of outermost oxygen atoms.[60]

**Non-bonded Potential.** Within MARTINIFF, a Lennard-Jones (LJ) 12-6 potential is used to model non-bonded interactions. The strength of interaction for a given pair of chemical building blocks, as given by LJ well-depth $\varepsilon_{i,j}$, is determined by calibrating against experimental thermodynamic data, such as oil/water partitioning coefficients. Overall the CG beads types are categorized as polar (P), non-polar (N), apolar (C), and charged (Q). The oil/water partitioning coefficients are not available for the clay particle, and role of high covalent coordination in significantly lowering the dispersion interactions[46] precludes use of individual building blocks of clay particle for determination of partition coefficients from AA simulations. We have instead used clay basal spacing and tactoid cleavage energy, experimental measurements are available for both, to parametrize $\varepsilon_{i,j}$ for CG clay beads.

The cleavage free energy (or surface tension) was obtained by considering a slow equilibrium cleavage process allowing for full surface reconstruction of clay particle. In this case, the entropic contribution to cleavage free energy is expected to be negligible for inter-gallery cations that do not consist of large alkyl tails, such as $Na^+$ and $TMA^+$.[61] We have therefore used the difference in total potential energy between two configurations (see Figure S2) — (state i) two clay particles in contact and (state ii) two clay particles separated by a large distance (10 nm), as an estimate of cleavage energy for the (001) surface. To mimic slow equilibrium cleavage process, the separated configuration was prepared by starting from the equilibrium configuration for two sheets in contact and moving one sheet perpendicular to (001) surface (i.e., along $z$) in steps of 0.5 nm, equilibrating each intermediate configuration for 2 ns. The simulation with two clay particles in contact is also used to determine basal spacing for the given clay model. For both the contact and separated (by 10 nm) configurations, a 1 ns simulation was carried out with semi-isotropic pressure coupling (along $x$ and $y$) and last 500 ps was used for determination of average potential energy. Further, analogous to classical surface theory,[62] we divided the cleavage energy into polar (electrostatic) and dispersive (Lifschitz-van der Waals) contributions.

We have used two model clays for identification of appropriate MARTINI types for CG beads. We obtained a reference level for interaction strength $\varepsilon_{i,j}$ by using a Pyrophyllite clay that has no isomorphic substitutions, and as a result, its cleavage energy is primarily composed of dispersive contribution only.[46,63] The [SiO$_2$] and [AlO$_2$H] units of pyrophyllite were modeled as SN0 type, representing a non-polar bead with no hydrogen-bonding ability. This bead type has been used for modeling non-polar silica surface.[64] Since, pyrophyllite cleavage energy is significantly overestimated with SN0 bead type, we have introduced a new non-polar bead type LN0 with reduced $\varepsilon_{LN0,LN0}$ to represent non-polar blocks in a clay particle (see Results and Discussion for details). Introduction of isomorphic substitutions in the octahedral layer has been shown to increase the polarity of surface SiO$_2$ blocks.[65] The following choice of TMA-MMT bead types led to an excellent reproduction of both polar and dispersive contributions to cleavage energy obtained in AA simulations: LNa (non-polar h-bond acceptor) for [SiO$_2$], LN0 (non-polar) for [AlO$_2$H], and LQ0 (charged) for [MgO$_2$H], These L-type beads with reduced dispersive interaction (w.r.t. MARTINI S-type), as found suitable to model high covalent coordination in a clay particle, are analogous to SNa, SN0, and SQ0 types, respectively, in MARTINIFF. The [TMA]$^+$ ion was modeled as MARTINI Q0 bead. The LJ size parameter $\sigma_{i,j}$ for any L-type bead is taken to be the same as that for MARTINI S-type beads, i.e., $\sigma_{i,j} = 0.43$ nm when both are smaller, finegrained (S or L) beads and $\sigma_{i,j} = 0.47$ nm otherwise. This default choice for $\sigma_{i,j}$ correctly reproduced the basal spacing for coarse-grained models of both pyrophyllite and TMA-MMT.

The procedure employed here for determination of LJ well-depth for L-type beads, and division of non-bonded LJ interaction energy (in MARTINIFF, well-depth $\varepsilon_{i,j}$ is dependent on polarity of building-blocks) into dispersive and polar components is described below briefly (full details in SI section Text S1). We consider the interaction strength between two non-polar beads, $\epsilon_{SN0,SN0}$ or $\epsilon_{LN0,LN0}$ (for L type), as purely dispersive. On this basis, the polar component of LJ interactions in MARTINIFF was defined as per equation 2.

$$\varepsilon_{i,j}^{polar} = \varepsilon_{i,j} - \varepsilon_{SN0,SN0} \ \forall \ \text{S-type or} \ \varepsilon_{i,j}^{polar} = \varepsilon_{i,j} - \varepsilon_{LN0,LN0} \ \forall \ \text{L-type} \quad (2)$$

We then defined the LJ well-depth for a pair of L-type beads such that the polar component of LJ interaction for L-type bead pair is same as that of S-type bead pair and the dispersive component is equal to $\varepsilon_{LN0,LN0}$ (equation 3).

$$\varepsilon_{i,j}^{(L)} = \varepsilon_{LN0,LN0} + \left(\varepsilon_{i,j}^{(S)} - \varepsilon_{SN0,SN0}\right) \quad (3)$$

The LJ well-depth for cross-interaction between a L-type bead with any MARTINI bead (not L-type) are taken to be the same as that for corresponding S-type bead with any MARTINI bead, for example, $\varepsilon_{LNa,SNa} = \varepsilon_{SNa,SNa}$, $\varepsilon_{LNa,Na} = \varepsilon_{SNa,Na}$, etc. The interaction matrix for LJ well-depths, incorporating the L-type beads and all other MARTINI beads used in the present study, is summarized in Table S2.

**2.2 MD simulation of Polymer-Clay System**

A single periodic-sheet of TMA-MMT was placed in the $xy$ plane with polymer molecules placed above the (001) basal surface. This arrangement mimics an exfoliated nanocomposite for large box dimension above (001) surface ($z$-axis) as used here ($\approx 4R_g$; see Figure S3). The clay sheet consisted of 960 unit cells ($20 \times 48 \times 1$) with an in-box size of 10.34 nm × 21.28 nm in $xy$. Three different polymers were used, namely polyethylene (PE80: 2240 $\frac{gm}{mol}$), polypropylene (PP100: 4200 $\frac{gm}{mol}$), and atactic polystyrene (PS60: 10400 $\frac{gm}{mol}$). The molecular weight was chosen to be just below the entanglement length of each polymer[48] so that equilibrium trajectories are sampled in reasonable simulation time. CHARMM36 FF is used for the polymers and [TMA]$^+$, and INTERFACE FF for the clay particle. Heinz et al.,[46] have shown that INTERFACE FF combined with CHARMM36 FF provided excellent reproduction of experimental data on various properties (cleavage energy, basal spacing, etc.) of Na-MMT and C18-MMT. For CG simulations, MARTINIFF parameters for polymer beads were taken from Rossi et al.[29] (PS) and Panizon et al.[28] (PE and PP). The LJ well-depth for all pairs of polymer-clay CG beads are given in Table S2. For each polymer, simulation was done at 1 bar and two temperatures above the corresponding glass transition temperature. Details of polymer-clay systems simulated here are given in Table 1. Slow polymer chain configuration relaxation and slow conformational dynamics in polymer melt systems necessitate a careful simulation set-up protocol. Some key details of the procedure used for obtaining equilibrium properties of polymer-clay system are given below with additional details in SI section Text S2.

**Preparation of simulation box at target density.** Equilibrium configuration of [TMA]$^+$ ions on the clay (001) surface was obtained using the same procedure as described in section 2.1. A diverse set of representative polymer chain configurations was taken from the equilibrium trajectory obtained for a polymer melt of 30 chains. Required number of chains were put above the clay particle (in Packmol$^{51}$) and box density was increased in several z-compression (5%) – energy minimization (steepest descent EM) steps till EM converged to a force less than 1000 $\frac{kJ}{mol}$. The box temperature was increased from 0 K to target temperature ($T_0$) in a 400 ps simulated annealing heating. Finally, a 20 ns simulation with anisotropic pressure coupling at 1 bar and periodic temperature annealing (each cycle 2 ns) between $T_0$ and $T_0 + 100$ K was used to obtain an equilibrated configuration. The average mean-squared displacement (MSD) of polymer center-of-mass (COM) reached up to $4R_g$ during periodic annealing. The polymer density and $R_g$ changed less than 2% in last 5 cycles. A similar protocol was used for preparation of an equilibrated configuration for respective polymer melt systems (without clay).

**Property calculation for polymer-clay systems.** Periodic annealing simulation was followed by a long simulation coupled to V-rescale temperature at $T_0$ and Parrinello-Rahman barostat at 1 bar. All properties were calculated from four repeat simulations (two sets of polymer chain configurations and two velocity seeds). Each simulation was run for $2\tau_B$, where Brownian relaxation time[66] ($\tau_B$) is defined as the time in which the chain MSD equals $R_g^2$ (Table 1). Separate averages are calculated for near and far regions, with near region defined as the region where significant layering of monomer density profile (along the normal to clay surface) is observed. In any given frame, a chain is assigned to the region which contains center of mass of 70% or more Kuhn segments. A chain remains unassigned if it does not meet this criterion for either region (more details in SI Section Text S2). A Kuhn length of 1.36 nm, 1.12 nm, and 2.95 nm was used for PE, PP, and PS (taken independent of molecular weight[48]), giving 13, 20, and 5 Kuhn segments in one chain, respectively.

**Table 1. Polymer-clay systems used for property determination**

| Polymer | Molecular Weight | Number of Chains (AA/CG) | Target temperature, $T_0$ (K) | $\tau_B$ (ns) | $t_S$ (ns) |
|---|---|---|---|---|---|
| PE80 | 2240 | 260/260 | 390 | 32 | 256 |
|  |  |  | 450 | 15 | 120 |
| PP100 | 4200 | 130/160 | 450 | 33 | 264 |
|  |  |  | 550 | 9 | 72 |
| PS60 | 10400 | 105/105 | 560 | 100 | 800 |
|  |  |  | 650 | 20 | 160 |

$\tau_B$= Brownian Relaxation time, $t_S$= total simulation time

***Two-body excess entropy.*** The two-body (pair correlation) contribution to conformational excess entropy, $s_2$, was calculated from the total radial distribution function (RDF) for polymer heavy atoms, $G(r)$. It has been shown that $s_2$ accounts for 80%–90% of the excess entropy for a Lennard-Jones fluid[67]. An ensemble invariant expression for $s_2$ of a fluid at density $\rho$ and temperature $T$ is given by equation (4).[68] For simulation in NPT ensemble, average number density of polymer heavy atoms, $\langle\rho\rangle$ was used.

$$s_{ex} \approx s_2 = -2\pi\rho \int_0^\infty [G(r)\ln G(r) - G(r) + 1]r^2 dr \qquad (4)$$

The monomer RDF was calculated separately in near and far regions by defining thin slices of width 0.37 nm along $z$. The *gmx rdf* utility of GROMACS was suitably modified to calculate RDF in a thin slice. An average for $s_2$ in near region was calculated from the value in 3, 3, and 2 slices in the near region and 4, 4, and 4 slices in the far region for PE, PP, and PS, respectively.

***Polymer configurational entropy.*** The chain configurational entropy, $s_c$, was calculated using the heuristic formula introduced by Schlitter which has been extensively tested and the expected error is within 5% for condensed phase systems.[69] Translational least-square fit was used to remove slowly converging degrees of freedom from the calculations, and therefore, we obtain the configurational entropy, instead of the absolute entropy.

***Polymer segment diffusivity.*** The diffusion coefficient ($D$) was calculated from the long-time slope of mean squared displacement (MSD) in $xy$, as given by the Einstein diffusion equation:

$$D = \frac{1}{4} \lim_{t\to\infty} \frac{d}{dt} \langle |r(t) - r(0)|^2 \rangle \qquad (5)$$

where, $\langle |r(t) - r(0)|^2 \rangle$ is the mean square displacement (MSD) of the particle center of mass. We have used the trajectory of center of mass of each Kuhn segment for determination of MSD. For calculation of $D$ in the near (far) region, trajectory data for a segment is used only for the time it is in the near (far) region and belongs to a chain assigned to the near (far) region. The linear part of MSD versus time data for calculation of slope was identified for each case. The MSD in this linear region ranged between $ml_b^2 - nl_b^2$, where $l_b$ is the Kuhn length, $m$ is 1, 1.7, and 1 for PE, PP, and PS respectively and $n$ is 4.5, 4.6, and 2.2 for PE, PP and PS respectively. The reduced diffusivity $D_R^*$, as defined by Rosenfeld[70,71] for characterizing thermodynamic-dynamic scaling in simple and complex fluids was also calculated, as per equation (6).

$$D_R^* = D \frac{\rho^{1/3}}{(k_B T/m)^{1/2}} \tag{6}$$

$D$ is scaled by macroscopic reduction parameters (density and temperature) defined in term of a mean inter particle distance, $d = \rho^{-1/3}$, and the thermal velocity, $v = (k_B T/m)^{1/2}$.

## 3. Results and Discussion

The self-consistency of developed MARTINIFF parameters for TMA-MMT – polymer system was verified by comparing estimates for select structural, thermodynamic, and dynamic properties obtained in all-atomistic simulations with that obtained in coarse-grained simulations. We have calculated all properties separately in near-clay and far-clay regions, as the former is expected to provide a stringent check for suitability of developed CG parameters for modeling the clay-polymer system. We have determined the influence of clay particle on properties of three polymer melts (PE, PP, and PS) at two temperatures to establish transferability of the developed parameters.

### 3.1 CG parameters for self-interaction of TMA-MMT particle

**CG bond and angle potential.** Elastic moduli of individual clay platelets[34] is required. Young's modulus $(E_x, E_y)$ was calculated from the slope of stress-strain curve obtained in the limit of slow uniaxial extension or compression of a periodic clay particle with no solvent (Figure S4 shows typical stress-strain curve obtained in AA and CG simulations). The Poisson ratio ($\nu_{yx}, \nu_{xy}$) and bending stiffness ($D_x, D_y$) were also determined from same set of simulations. Table 2 lists various property values obtained for a TMA-MMT particle at 298 K. For all calculations, we have used a clay thickness of $h = 0.937$ nm, defined as the distance between center of surface oxygen atoms in two tetrahedral layers in addition with diameter of oxygen atom. We obtained $E_x$ ($E_y$) of 416 GPa (418 GPa) and $\nu_{yx}(\nu_{xy})$ of 0.151 (0.154) from AA simulations of a single clay sheet. The obtained Poisson ratio compares well with experimentally determined value of 0.22-0.24 for corresponding bulk clay aluminosilicates.[72] However, single sheet Young's modulus cannot be compared to experimentally determined value for corresponding bulk mineral as individual sheets are held together only by weaker physical interactions (in contrast to chemical bonds in a single sheet) and the intergallery space is less dense than the intrasheet sheet layers. Stishovite provides for a good comparison as it has a three dimensional bonded network based on polymerization of octahedra and has no intergallery space $\left(\text{density } 4300 \ \frac{\text{kg}}{\text{m}^3}\right)$, and its $E$ of 453 GPa[73] compares well with the value obtained for single sheet. We note that the mechanical properties have a relatively strong dependence on choice of force field, with $E_x$ ($E_y$) of 320 GPa (344 GPa) and $\nu_{yx}(\nu_{xy})$ of 0.44 (0.49) obtained with CLAYFF force field.[74]

The above reported mechanical properties were reproduced for the coarse-grained sheet by tuning the stiffness of harmonic bond and angle potentials, $k_b^{CG}$ and $k_\theta^{CG}$, respectively. To correctly represent high stiffness of a clay particle we expect both values to be much higher than typical MARTINIFF values for bond and angle potentials. An initial guess for $k_b^{CG}$ was obtained by mapping a CG bond potential on sum of all bonded potentials between two CG beads, viz. two AA bond and one AA angles. Also, as an initial guess, $k_\theta^{CG}$ was taken equal to $k_\theta^{AA}$ on account of space-frame like structure of clay sheet. Since bond stiffness is expected to have a greater impact on $E$, $k_b^{CG}$ was varied at fixed $k_\theta^{CG}$ till the sum of relative error in $E$ and relative error in $D$ is minimized. This gave $k_b^{CG} = 179912 \frac{\text{kJ}}{\text{mol}} \left(= 0.5 k_b^{AA}\right)$ and $k_\theta^{CG} = 1422 \frac{\text{kJ}}{\text{mol}} \left(= k_\theta^{AA}\right)$, at which both Young's modulus and bending stiffness of the CG sheet compared excellently with the corresponding AA sheet values (Table 2). We recognize that other combinations of $k_b^{CG}$ and $k_\theta^{CG}$ might yield a similar favorable comparison.

**Table 2.** Elastic properties $(E, \nu, D)$ for TMA-MMT obtained by uniaxial extension or compression of a single sheet in all-atomistic (AA) and coarse-grained (CG) simulations. Cleavage free energy $(\sigma)$ and its break-up into dispersive $(\sigma_d)$ and polar contributions $(\sigma_p)$, and basal spacing $(d)$ for pyrophyllite and TMA-MMT calculated from simulations on two identical sheets.

| | AA | CG |
|---|---|---|
| $E_x, E_y$ (GPa) | 416, 418 | 407, 408 |
| $\nu_{yx}, \nu_{xy}$ | 0.154, 0.151 | 0.172, 0.171 |
| $\frac{D_x}{h^2}, \frac{D_y}{h^2}$ (N/m) | 34.29, 34.54 | 35.01, 35.03 |
| Pyrophyllite $\sigma$ (mJ/m²) | 40.1 | 40.2 |
| $\sigma_d, \sigma_p$ (mJ/m²) | 32.1, 8.0 | 32.0, 8.2 |
| $d$ (nm) | 0.92 | 0.916 |
| TMA-MMT $\sigma$ (mJ/m²) | -105.6 | -115.1 |
| $\sigma_d, \sigma_p$ (mJ/m²) | -20.37, -85.2 | -21.2, -93.8 |
| $d$ (nm) | 1.31 | 1.299 |

**CG nonbonded potential.** The self-interaction between clay beads, quantified using the cleavage free energy for the (001) basal surface $(\sigma)$ and the basal spacing $(d)$, was used to determine appropriate MARTINI type for various CG beads in TMA-MMT. As explained in methods section, the LJ size parameter $\sigma_{i,j}$ was taken to be the same as that for finegrained S-type ring beads defined in MARTINIFF and this gave correct basal spacing for both pyrophyllite and TMA-MMT (Table 2). Further, since MARTINI bead types are categorized on basis of polarity of the chemical block, we calculated separate polar (electrostatic) and dispersive (Lifschitz-van der Waals) contributions to cleavage free energy. The LJ well-depth for CG beads, $\varepsilon_{i,j}$, was accordingly divided into a polar and dispersive contribution as per equation 2. We first selected an existing MARTINI bead type SN0 (non-polar with no h-bonding ability) for both [$SiO_2$] and [$AlO_2H$] units of pyrophyllite, a phyllosilicate with no isomorphic substitutions resulting in a cleavage energy primarily composed of dispersive contribution only.[46,63] This bead type has been used for modeling non-polar silica surface.[64] We obtained a cleavage energy of $-150 \frac{mJ}{m^2}$ for coarse-grained pyrophyllite as against a value of $-40 \frac{mJ}{m^2}$ from AA simulation. Several other studies have reported an overestimation of vdW interactions as well as overall interaction (quantified in terms of second virial coefficients) on use of MARTINIFF, for example, in hydration of organic compounds,[40] protein-protein binding,[75] and aggregation of polysaccharides.[76] This effect is further accentuated in the present case because of dense covalent bonding and significant ionization in phyllosilicates, which were shown to lead to distinctly lower well depths for LJ dispersion interaction in the INTERFACE FF for sheet atoms in comparison with corresponding rare gases.[46] Not surprisingly, we found that a simple scaling down of $\varepsilon_{i,i}$ of SN0 bead leads to an excellent reproduction of corresponding AA values for $\sigma$ and $d$ of pyrophyllite (Table 2). This bead with a reduced $\varepsilon_{i,i}$ ($= 0.213\varepsilon_{SN0,SN0}$) is referred to as LN0, a new non-polar finegrained bead found suitable for modeling the dispersive contribution to surface energy of phyllosilicates which have a dense covalent bonding framework.

The cleavage free energy for TMA-MMT is expected to provide a more stringent test because of larger diversity in polarity of constituent chemical blocks, represented by [$SiO_2$], [$AlO_2H$], [$MgO_2H$]$^-$, and [$TMA$]$^+$. The octahedral layer blocks, [$AlO_2H$] and [$MgO_2H$]$^-$ were modeled as LN0 (non-polar) and LQ0 (charged), respectively. The surface tetrahedral layer [SiO$_2$] block was modeled as LNa (non-polar h-bond acceptor), since introduction of isomorphic Al$^{3+}$ → Mg$^{2+}$ substitutions in the octahedral layer has been shown to increase the polarity of surface [SiO$_2$] blocks.[65] The dense covalent network is expected to suppress only the dispersive interactions, and therefore, the polar contribution to LJ well-depth ($\varepsilon_{i,j}$) for L-type beads is taken to be the same for MARTINI S-type beads. Accordingly, $\varepsilon_{i,j}$ for self- and cross-interaction between various L-type beads is defined as per equation 3. These choices gave an excellent agreement for cleavage energy as well as its polar and dispersive components calculated from CG and AA simulations (Table 2). We note that the cleavage free energy, σ of $-105.6 \frac{mJ}{m^2}$ obtained for the TMA-MMT sheet in AA simulation is, as expected, larger than the value of $-140 \frac{mJ}{m^2}$ obtained for Na-MMT[46] using the Interface FF. This difference can be explained on basis of larger basal spacing for the former—1.31 nm for TMA-MMT in contrast to 1.08 nm for Na-MMT. Taking the distance scaling of columbic potential energy, which dominates the contribution to σ, the cleavage energy of Na-MMT will increase to $-108 \frac{mJ}{m^2}$ at a separation of 1.31 nm.

Ricci et al.[77] showed that when their variational implicit solvent method (VISM) is combined with MARTINIFF, a simple downscaling of MARTINI $\varepsilon_{i,j}$ values bought solvation energies as well as its partitioning among coupled energy terms (hydrophobic, vdW, and electrostatics) from coarse-grained approach to a good quantitative agreement with the ones from atomistic approach. We note that in all the previous cases MARTINI $\varepsilon_{i,j}$ were downscaled uniformly, while in present case we find that partitioning each $\varepsilon_{i,j}$ into polar and dispersive components and downscaling only the dispersive component provides best agreement for partitioning of TMA-MMT cleavage energy among the polar and dispersive contributions.

## 3.2 Structural, thermodynamic, and dynamical properties of clay-polymer systems.

A model exfoliated polymer-clay nanocomposite was prepared by adding a thick layer of polymer melt ($\approx 4R_g$ along z-axis) above the basal surface of clay (see Figure S3). The three polymer systems investigated here are PE80 (80 ethylene monomers), PP100 (100 propylene monomers), and PS60 (60 styrene monomers). Some key differences between the chosen polymers include use of normal MARTINI bead (two ethylene monomers) for PE versus finegrained S-type bead (one propylene monomer) for PP, with both coarse-grained representations consisting of only backbone beads. In contrast, PS consists of a large side-chain (phenyl group) modeled by a ring of three S-type MARTINI beads. Below, we report data on comparison of an extensive set of properties determined from the coarse-grained approach versus the atomistic approach. For all three polymers, simulations were carried out at a lower temperature around 50 K above respective glass transition temperatures $(T_g)$ and a higher temperature around 100-150 K above $T_g$.

**Polymer density profile normal to clay surface.** We first validate simulation equilibration by comparing average density obtained here for polymer melts (without clay particle), $\rho_0$, with that in literature. The melt properties are also used a reference level for determining effect of clay on polymer. The $\rho_0$ obtained from the AA simulations are $780 \frac{\text{kg}}{\text{m}^3}$, $728 \frac{\text{kg}}{\text{m}^3}$, and $890 \frac{\text{kg}}{\text{m}^3}$ for PE80 at 390 K, PP100 at 450 K, and PS60 at 560 K, respectively. The corresponding values from CG simulations are $790 \frac{\text{kg}}{\text{m}^3}$, $1039 \frac{\text{kg}}{\text{m}^3}$, and $925 \frac{\text{kg}}{\text{m}^3}$, respectively. The densities of coarse-grained polyethylene and polystyrene melts are in excellent quantitative agreement with the ones from AA simulations. The much larger estimate for $\rho_0$ obtained in MARTINI CG model for polypropylene melt is in accordance with that reported by Panizon et al.,[28] whose CG parameters are taken in this study. It was reported that this choice of MARTINI CG parameters, wherein a propylene monomer is represented by a finegrained S-type bead, gives $R_g$ and $R_e$ values in excellent agreement with values obtained from atomistic simulations in the melt as well as in a good solvent (1,2,4-trichlorobenzene). A larger bead diameter decreased the CG melt density, however, lead to significant differences in chain conformations. The bulk densities as calculated in our AA simulations compare excellently with experimental measurements as well as simulation data on low molecular weight polymers. The density of PE36 melt (measured by a specific gravity bottle) was reported to be $763 \frac{kg}{m^3}$ at 450 K[78] and the density of PP100 melt (measured by a dilatometer) was reported to be 710-764 $\frac{kg}{m^3}$ at 450 K.[79] Density of a united-atom model of PE26 melt and MARTINI CG model of PE20 melt was reported to be $750 \frac{kg}{m^3}$ and $760 \frac{kg}{m^3}$, respectively (both at 450 K).[28] Rossi et al.[29], whose parameters for polystyrene are used here, reported a density of $951 \frac{kg}{m^3}$ and $938 \frac{kg}{m^3}$ at 500 K for AA and CG model of PS30 melt, respectively. Overall these results indicate correct determination of equilibrium densities of polymer melts in this study.

All three polymers show layering near the clay surface with width of near-clay region defined as the distance up to which significant layering is observed (Figure 2). In all cases except for CG polystyrene, the densities in the far region reach the value for the corresponding polymer melt. For CG clay-polystyrene system, a density of $1.05\rho_0$ is obtained in the far regime. For all three polymers, at both temperatures, the monomer density profiles obtained from CG and AA simulations show a fairly good agreement. The observed broadening of peaks in AA simulations is due to the use of center of mass of atoms in corresponding CG beads in determination of monomer density profiles.

Several computational studies have shown that polymer chains near solid surfaces are highly affected with respect to the isotropic bulk wherein polymer-clay interactions govern overall structural and dynamic behavior in the near surface regime.[80] A density peak of 3-5 times the bulk value was reported for strongly attractive interactions in case of polyethylene near a graphite sheet[4] and polyethylene oxide near Na-MMT.[81] In contrast, we observe the first peak height for polymers near TMA-MMT surface to be around 1.5 times the bulk value, indicative of weakly attractive interactions. The effect of clay on the monomer density profile is observed up to a distance of approximately 1.5 nm ($\sim R_g$) form the clay surface, as also observed for the PE-graphite[4] and PS–silica nanoparticle systems,[82] indicating a smaller effect of polymer-clay interactions on the range of clay induced layering. We have therefore used the layered monomer density profiles to divide the simulation box in two regions: (i) the near-clay region (from the clay surface up to the third peak of density profile) and (ii) the far region (at distances greater than that of the third peak). For all cases the monomer density variations are not significant beyond the third peak, and hence, polymer properties are expected to be significantly altered by the clay only in the as-defined near region. Small shifts in this boundary between near and far regions does not affect our conclusions.

**Effect of clay on polymer chain conformation.** We have characterized polymer chain conformation using radius of gyration ($R_g$) and end-to-end distance ($R_e$), with average and standard deviation used to define the distribution. The data for all three polymer-clay systems as well as polymer melts is reported in Table 3. The $R_g$ values obtained from AA simulations compare fairly well with reported experimental and simulation data— 2.12 nm for PE80 melt at 390 K versus 2.38 nm for PE88 at 450 K obtained by Panizon et al.[28] using configurational-bias Monte Carlo scheme developed by Smit et al.[83], 1.95 nm for PP100 melt at 450 K versus 2.20 nm for PP100 obtained using Small-angle neutron scattering studies[84], and 1.95 nm for PS60 melt at 560 K versus 2.01 nm for PS60 obtained using Small-angle X-ray scattering studies.[85]

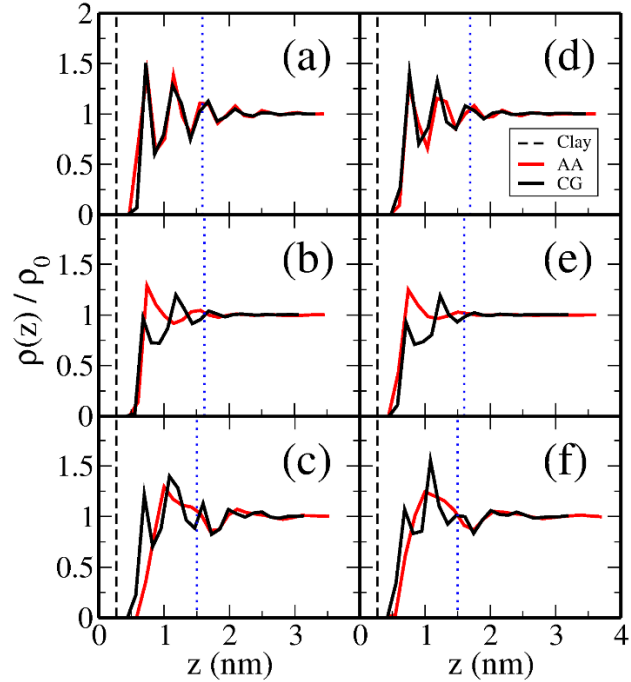

**Figure 2.** Density profile of monomers normal to clay surface. The density values are reported as $\rho(z)/\rho_0$, where $\rho_0$ is the corresponding value for respective melts. (a-c) Normalized monomer density profile for PE, PP and PS at the lower temperature value of T = 390 K, 450 K and 560 K, respectively and (d-f) at the higher temperature value of T = 450 K, 550 K and 650 K respectively. The location of center of surface Si atoms is shown by black dashed line. The boundary of near region (up to third peak in density profile), shown by blue dotted line, is at 1.33 nm, 1.33 nm, and 1.23 nm from surface Si atoms, for PE, PP, and PS, respectively.

In all cases, the $R_g$ and $R_e$ values are higher in the near region and essentially the same in the far region w.r.t. the values in the melt. The ratio of $R_g$ values in the near and far regions, a measure of effect of clay on the polymer conformation, is 1.06, 1.08, and 1.14 for the PE80, PP100, and PS60 chains, respectively, in the AA simulations. The corresponding ratio in CG simulations is 1.07, 1.04, and 1.15, respectively, indicating that extent of increase in $R_g$ near the clay surface is quantitatively reproduced by the CG parameters. Previous studies have ascribed this increase in polymer chain size near a surface to a combination of polymer-clay interaction as well as entropic chain re-orientation effects. The latter was used to explain the observed increase in component of $R_g$ parallel to a hard wall wherein there are no enthalpic interactions between the clay and polymer beads.[86] Strongly attractive surface - polymer interactions in case of polyethylene oxide near Na-MMT were shown to lead to up to two times (w.r.t. melt) increase in $R_g$ of near surface polymer chains.[81,87] Cho et al.[88] reported elongation of polyethylene chains (higher $R_e$) near the solid surface and linked it to energetically favorable polymer-surface interactions. Similarly, neutron scattering data on ultra-thin films (12 nm) of polystyrene on glass substrate revealed an increase of up to 1.5 times in the parallel component of $R_g$.[86] In contrast, repulsive surface - polymer interactions were shown to lead to a slight decrease in $R_g$ of near surface chains. The increase in $R_g$ in the range of 1.06-1.14 for the systems in this study is indicative of weakly attractive surface-polymer interactions for PE80, PP100, and PS60 chains on TMA-MMT. Therefore, the chain conformation data is in sync with observation of weak layering in monomer density profiles perpendicular to clay surface.

**Table 3. Polymer chain structural properties.** Radius of gyration and end-to-end distance of polymer chains in the near and far regimes as well as polymer melt for polyethylene at 390 K, polypropylene at 450 K, and polystyrene at 560 K. Average and standard deviation was calculated from 4 independent simulations for each system.

| Polymer | | $<R_g>$ (nm) | | $<R_e>$ (nm) | |
|---|---|---|---|---|---|
| | | AA | CG | AA | CG |
| PE80 | Near | 2.21 ± 0.24 | 2.31 ± 0.29 | 5.28 ± 0.42 | 5.05 ± 0.44 |
| | Far | 2.07 ± 0.24 | 2.16 ± 0.26 | 4.87 ± 0.40 | 4.66 ± 0.39 |
| | Melt | 2.12 ± 0.20 | 2.18 ± 0.16 | 5.02 ± 0.38 | 4.72 ± 0.36 |
| | | 2.38[a] | 2.21[b] | | |
| PP100 | Near | 2.128 ± 0.18 | 2.28 ± 0.21 | 4.88 ± 0.43 | 4.77 ± 0.50 |
| | Far | 1.967 ± 0.17 | 2.19 ± 0.22 | 4.61 ± 0.41 | 4.49 ± 0.47 |
| | Melt | 1.955 ± 0.18 | 2.20 ± 0.18 | 4.62 ± 0.42 | 4.48 ± 0.46 |
| | | 2.20[c] (expt.) | 2.22[d] | | |
| PS60 | Near | 2.184 ± 0.11 | 2.38 ± 0.18 | 5.42 ± 0.35 | 5.81 ± 0.45 |
| | Far | 1.918 ± 0.10 | 2.08 ± 0.14 | 4.96 ± 0.35 | 5.27 ± 0.44 |
| | Melt | 1.950 ± 0.11 | 2.10 ± 0.21 | 4.92 ± 0.31 | 5.32 ± 0.44 |
| | | 2.01[e] | 2.12[f] | | 5.40[g] |

[a] PE88 simulated using configurational-bias Monte Carlo scheme developed by Smit et al.[83], [b] PE80 (40 CG beads) melt simulated at 450 K by Panizon et al.[28], [c] PP100 (experimental value obtained by fitting Small-angle neutron scattering (SANS)[84] data, [d] PP100 melt simulated at 450 K by Panizon et al.[28], [e] PS60 ( experimental value obtained using Small-angle X-ray scattering studies[85]) . [f & g] PS60 (CG simulation by Rossi et al.[29].

**Radial distribution function and conformational excess entropy.** Layering of fluid near a surface or in confinement is typically not a direct measure of effect of surface on thermodynamic and dynamical properties of fluid. Karatrantos et al.[88] observed no difference in the monomer-monomer RDFs for the polymer nanocomposite system and the pure melt even though the polymer density near the solid surface was higher than that in bulk. Using model wall-colloid interactions it was shown that inferences made on dynamical properties on the basis of extent of particle layering can be misleading and conformational excess entropy ($s_{ex}$) was instead used to quantitatively explain the link between colloidal particle diffusion coefficient and wall-colloid interactions.[89] To this end, we have calculated the two-body contribution to excess entropy, $s_2$, using bead-bead RDFs in thin slices parallel to the clay sheet. In fact, because of this quantitative link between RDFs and thermodynamic and dynamic properties, the RDFs are sometimes used as a reference property for determination of effective potentials between coarse-grained beads.[31]

We have calculated the 2D RDFs, $g_{bb}^{2D}(r)$, using either the center of CG bead, or in case of AA simulations, using the center of mass of atoms in corresponding CG bead. The comparison of RDFs obtained from the coarse-grained simulations with those from the atomistic simulations (shown in Figure 3) will provide a very stringent test, since, within the MARTINI approach the CG parameters are not obtained using RDF matching. Previous studies on MARTINIFF models of the polymers used here did not report comparison between atomistic and coarse-grained RDFs.[28,29] The bead-bead RDFs for PE80 – TMA-MMT system as obtained in CG simulations compares very well with that obtained in AA simulations in both near and far regions at both 390 K and 450 K. Again, the broadening and lowering of RDF peak for AA simulations can be mainly attributed to use of center of mass of atoms in a CG bead. In contrast, the RDFs for polypropylene and polystyrene obtained from coarse-grained and atomistic approach compare poorly, with key difference being an outward shift in the first peak of atomistic RDFs. The outward shift is particularly noticeable for polypropylene for which use of smaller, fine-grained MARTINI S-type bead provided a better agreement with atomistic $R_g$ and Re values but led to higher density for coarse-grained melt. We acknowledge that further work is required to obtain a better set of MARTINI parameters for modeling polypropylene melt. Finally, we note that for all cases there is a small increase in local ordering in the near-clay region which is clearly manifest by larger $|s_2|$ in the near region w.r.t. that in the far region / melt, and this difference in $s_2$ between near and far regions is correctly captured by CG simulations (Figure S5). We will re-visit this difference in $s_2$ in the near and far regions in our discussion on diffusion-entropy scaling and show that changes in diffusion coefficient near the clay surface are quantitatively linked to changes in $s_2$.

**Chain conformational entropy.** The differences in static fluctuations of a polymer chain in the near and far regions can be adequately characterized by chain conformational entropy. Therefore, it provides another quantitative measure (besides $R_g$ and $R_e$) for determining effect of clay on the polymer conformation and validate developed CG parameters. Figure 4 shows conformational entropy per bead ($S_c$) for each of the AA and CG clay-polymer systems. $S_c$ for the atomistic simulation was calculated using one atom per CG bead (see methods for details). The $S_c$ values for all three polymers, at both lower and higher temperature level, and in both near and far regions show an excellent quantitative agreement between AA and CG simulations with a maximum deviation of 1.5% (for PS60 in near-clay region). Further, as expected, $S_c$ is lower in the near region w.r.t. the value in the far region. At the lower temperature level, $S_c$ determined from atomistic simulation is smaller by 2.4, 3.5, and 5.4 $\frac{J}{mol.K}$ for PE80, PP100, and PS60, respectively. The corresponding decrease as calculated from the CG simulations are 2.2, 3.2, and 5.8 $\frac{J}{mol.K}$, respectively. This represents a decrease of 4%-11% in $S_c$ in the near region, with the difference between near and far regions increasing with increase in side-chain group size from polyethylene, to polypropylene to polystyrene. A similar trend is observed at the high temperature level. This effect of a surface on chain static fluctuations can be attributed to combined effect of surface-polymer interactions and polymer packing near a surface. In both CG and AA density profiles, we observe that the phenyl group of polystyrene is preferentially oriented towards the clay surface leading to relatively larger decrease in near-clay $S_c$. Similar decrease in conformational entropy induced by presence of a polymer chain near a surface has been reported for a variety of polymer chain models in different solvents. For confinement between parallel plates with slit-width $\approx R_g$, a decrease of 7.5% in $S_c$ was reported for a PE50 chain in an aqueous solution[90] and a decrease of 4.7% was reported for a 200-bead SAW chain in a SAW melt.[91] The CG models for polymer molecules used here were validated by comparing reproduction of property changes with change in solvent.[28,29] The transferability of CG model of clay-polymer system developed here needs to be established for the case of a significant change in solvent (for example, replacing polymer melt by and aqueous solution) and will be part of a future study. Finally, we note that extent of decrement in $S_c$ is fairly well reproduced in the coarse-grained model for polypropylene even though the CG model does not contain an explicit side-chain bead. This coincides with quantitative agreement for $R_g$ and $R_e$ values of polypropylene in CG and AA simulations even when the CG model gives a 42% larger melt density. Since any given coarse-graining techniques can use only a few properties to parameterize the CG forcefield, it is important to validate self-consistency of developed parameters by comparing a large set of distinct properties.

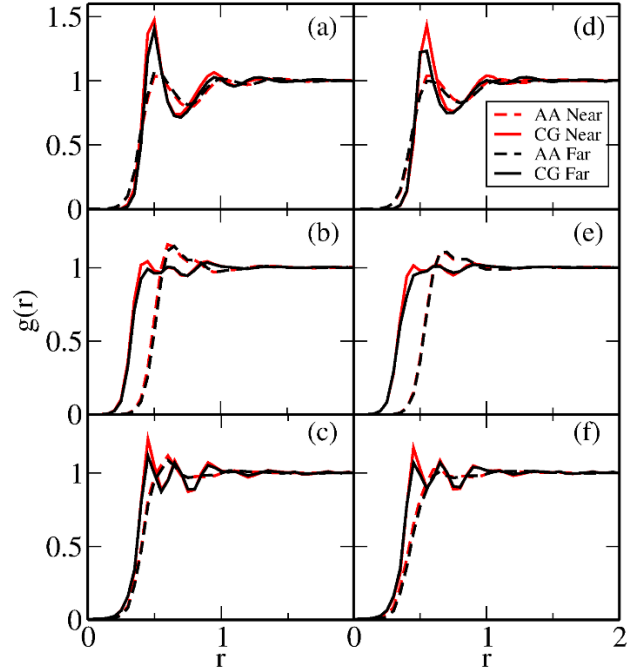

**Figure 3**. 2D radial distribution function for effective CG beads, $g_{bb}^{2D}(r)$. The 2D RDF was calculated in thin slices parallel to clay sheet and averaged separately for the near and far regions. Data is shown respectively for (a-c) PE80, PP100, and PS60 above the clay at lower temperature ($T = 390$ K, 450 K, and 560 K) and (d-f) at higher temperature ($T = 450$ K, 550 K and 650 K).

**Chain self-diffusion coefficient.** Since MARTINIFF has been parameterized using thermodynamic properties, comparison of dynamic properties between atomistic and coarse-grained approach will provide a self-consistency check for developed CG parameters. Here, we have used changes in self-diffusion coefficient of the polymer near TMA-MMT surface as one dynamic property for validation of CG parameters. We have used long trajectories, with total

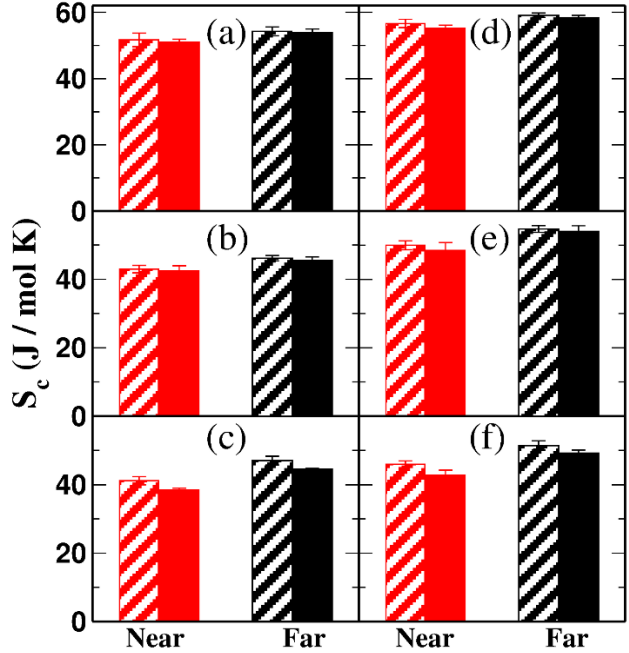

**Figure 4.** Chain conformational entropy per effective CG bead, $S_c$ averaged separately for the near and far regions. For the AA simulation, one atom per CG bead was taken for calculation of $S_c$. Data is shown respectively for (a-c) PE80, PP100, and PS60 above the clay at lower temperature ($T$ = 390 K, 450 K, and 560 K) and (d-f) at higher temperature ($T$ = 450 K, 550 K and 650 K). AA data is shown by textured bars and CG by solid bars.

simulation time large enough to obtain chain center of mass mean-squared displacement up to eight times $R_g^2$, as essential for accurate determination of diffusion coefficients. The self-diffusion coefficient for chain center of mass, $D_{\text{chain}}$, obtained here using the atomistic simulations compare fairly well with data reported using modified nuclear magnetic resonance (NMR) field gradient technique.[92] We obtained a $D_{\text{chain}}$ of $2.82 \times 10^{-7}$ and $7.01 \times 10^{-7} \frac{\text{cm}^2}{\text{s}}$ for PE80 melt at 390 K and 450 K, respectively. The corresponding value reported from NMR measurements on a PE88 melt is $2.2 \times 10^{-7}$ and $5.4 \times 10^{-7} \frac{\text{cm}^2}{\text{s}}$ at 400 K and 454 K, respectively. Similarly, a $D_{\text{chain}}$ of $6.2 \times 10^{-8} \frac{\text{cm}^2}{\text{s}}$ for
30

PS60 melt at 560 K obtained from atomistic simulations compares well with NMR estimate of 2.7x10$^{-8}$ $\frac{cm^2}{s}$ for PS53 melt at 500 K.

To determine the role of clay surface on polymer diffusion we have calculated the self-diffusion coefficient of effective segments (taken equal to Kuhn length), $D_s$, in the near-clay and far region separately (see methods for details). We note that the CG simulation timescale is expected to be different from atomistic simulation timescale.[29] For example, we obtain a value of 1.41 and 0.6 for the ratio of $D_{chain}$ from CG and AA simulations in PS60 at 560 K and 650 K respectively, while Rossi et al[29] reported this ratio to be 1.5 at 540 K. The observed decrease in diffusivity ratio of coarse-grained to atomistic with increase in temperature is also in accord with Rossi et al.[29] Therefore, we compare only the relative change in near and far region $D_s$ w.r.t. the value in melt, $D_{s0}$. Figure 5 shows this comparison of normalized self-diffusion coefficient for effective segments in all three polymer-clay systems at both temperature levels. As expected, $D_s/D_{s0}$ is almost unity in the far region for both PE80 and PP100 chains in all cases, and for PS60 chains in the atomistic simulation. Segmental diffusivity in far region of coarse-grained PS60-clay system is about 62.2% and 74.3% of the melt value at 560 K and 650 K, respectively. This discrepancy, which can be related to larger layering observed in the monomer density profiles in the CG model (Figure 3(c), (f)), points towards inadequacy of the CG parameters developed for the PS60-clay system.

For all three systems, both atomistic and coarse-grained simulations show a decrease in $D_s$ in the near region compared to that in the far region. The ratio of near region to far region $D_s$ obtained from atomistic simulation is 0.79, 0.73, and 0.34 for PE, PP and PS polymer clay systems, respectively, at the lower temperature level. The corresponding ratio from coarse-grained simulation is 0.75, 0.64, and 0.33, respectively (Figure 5(a)-(c)). The difference in near and far region $D_s$ decreases at higher temperature level ((Figure 5(d)-(f)). Therefore, the CG model quantitatively captures the slowdown in diffusion near the clay surface as well as its temperature dependence. The diffusion coefficient of polyethylene oxide chains near a silicate surface decreased by approximately half that of the remaining layers.[81] Similarly, a significant decrease in near surface chain diffusion compared to the bulk value was reported for strongly attractive interactions in case of polyethylene near a graphite surface.[4] In contrast, a small increase in chain diffusivity near a smooth surface was reported in case of repulsive polymer-surface interactions.[93] In the same study, a decrease in diffusivity in range of 0.6-0.9 was reported in case of attractive polymer-surface interactions. The extent of reduction in near surface $D_s$ for PE80-clay and PP100-clay systems is indicative of weakly attractive interactions and/or near surface chain re-orientation, analogous to similar observation of effect of TMA-MMT on monomer density profiles, $R_g$, $R_e$, and chain conformational entropy. The reduction in near surface $D_s$ is much larger for PS60-clay system, captured by both atomistic and coarse-grained models, the observed effect being even larger than that reported for case of strongly attractive polymer-surface interactions. Since polystyrene is hydrophobic and TMA-MMT is hydrophilic we believe this much stronger effect of surface on PS60 diffusion is linked to optimal near surface packing of polymers with a large side-chain group. We quantify this connection using diffusion-excess entropy scaling, where the latter captures static pair and higher order correlations between atom or bead positions.

### 3.3 Diffusion-entropy scaling

In the previous section we have shown a consistent effect of clay surface, temperature, and polymer type on polymer chain structure ($R_g$, $R_e$), polymer morphology near clay surface (monomer density profile, 2D RDFs), chain and effective segment diffusion coefficients, and conformational entropy. This structure-property relationship can be further quantified by employing Rosenfeld's[70,71] diffusion-excess entropy scaling which has been used to correlate dynamical properties with conformational excess entropy, a thermodynamic property which is directly obtained from pair and higher order static positional correlations. This diffusion-excess entropy scaling has been shown to be applicable for a variety of simple and complex fluids at a large range of pressure and temperature, as well as for fluids in confinement. We have used the two-body (pair correlation) contribution to excess entropy, $s_2$, obtained from 2D RDFs, defined as per equation 4 (see methods section). The near region and far region $s_2$ was calculated from respective RDFs. We have used all polymer CG beads for calculation of $s_2$ in the coarse-grained model and all polymer heavy atoms for the atomistic model. Figure 6(a),(b) show the data on reduced self-diffusion coefficient for effective segments, $D_R^*$, plotted versus $s_2$ for the atomistic and coarse-grained simulations, respectively. In both cases, the data corresponding to three polymers (PE80, PP100, PS60), two temperature levels, two systems (polymer melt and polymer-clay), and two regions (near-clay and far) overlap on essentially a single master curve. The data sets for atomistic and coarse-grained simulations are excellently fit by $D_R^* = 0.033 e^{-1.57 s_2}$ and $D_R^* = 0.28 e^{-4.85 s_2}$, respectively. Similar exponential scaling for diffusion-excess entropy has been obtained for a diverse set of systems.[70,71] Figure 7 shows the analogous scaling of $D_R^*$ with chain conformational entropy, $S_c$. An exponential scaling is observed in this case, albeit with a larger scatter in both atomistic and coarse-grained data. This shows that excess entropy which is based on static positional correlations of intra- and inter-chain atoms/beads provides a better probe for determining effect of surface on polymer dynamics than conformational entropy which is based on only intra-chain positional fluctuations, even though the latter will also be (indirectly) affected by the local environment of a given chain.

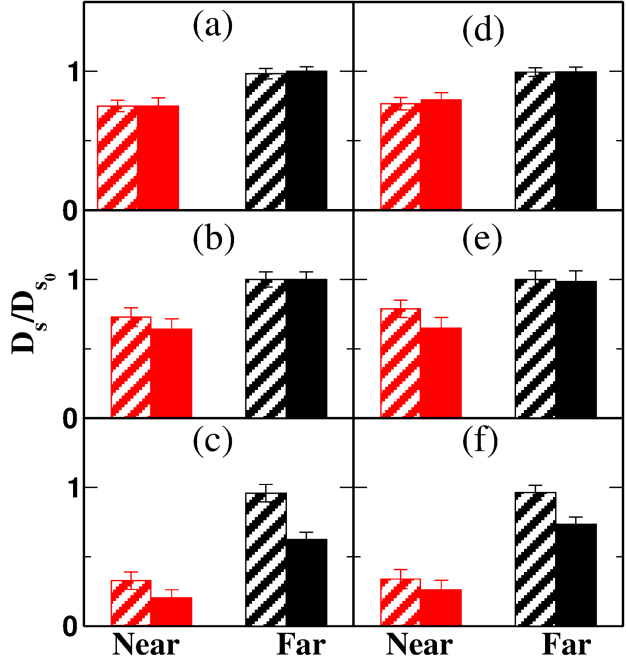

**Figure 5.** Normalized effective segment diffusivity, $D_s/D_{s_0}$, averaged separately for the near and far regions. The effective segment size was taken to be 6, 5, and 12 monomers for PE80, PP100, and PS60 chains, respectively. Data is shown respectively for (a-c) PE80, PP100, and PS60 above the clay at lower temperature ($T = 390$ K, 450 K, and 560 K) and (d-f) at higher temperature ($T = 450$ K, 550 K and 650 K). AA data is shown by textured bars and CG by solid bars.

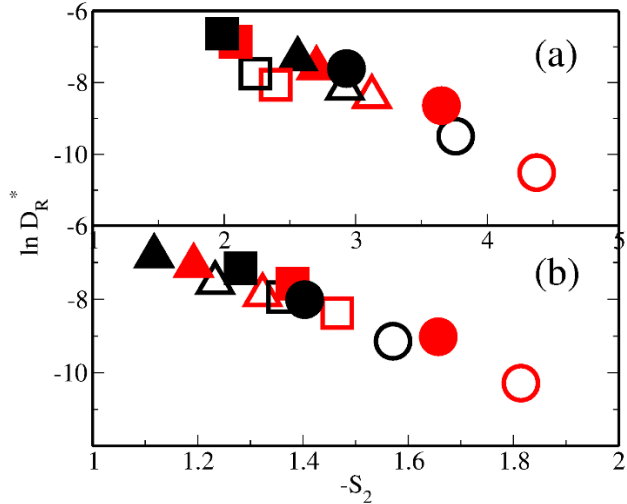

**Figure 6**. Variation in reduced effective segment diffusivity, $D_R^*$, plotted versus two-body contribution to conformational excess entropy for (a) atomistic simulations and (b) coarse-grained simulations. Symbol key: near region (red) and far region (black), low temperature level (open) and high temperature level (filled), PE80 (triangle), PP100 (square), and PS60 (circle).

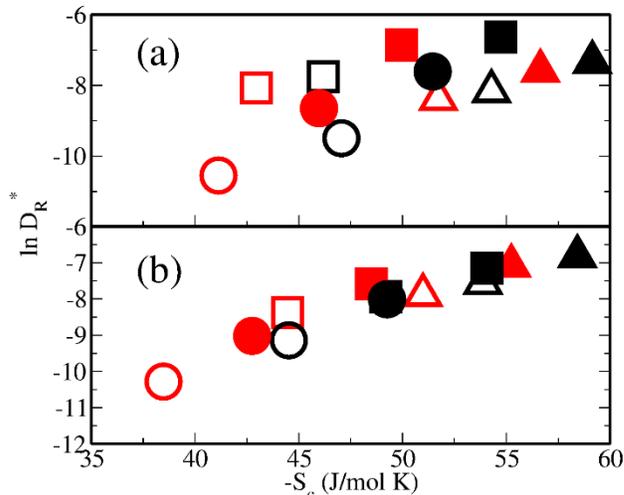

**Figure 7.** Variation in reduced effective segment diffusivity, $D_R^*$, plotted versus chain conformational entropy $S_c$ for (a) atomistic simulations and (b) coarse-grained simulations, respectively. Symbol key is same as figure 6.

## 4. Conclusions

We have reported development of a chemically specific coarse-grained model for a montmorillonite-polymer system based on the MARTINI forcefield. Montmorillonite (MMT) is an aluminosilicate clay mineral (phyllosilicate) with isomorphic $Al^{3+} \rightarrow Mg^{2+}$ substitutions in the octahedral layer. We have shown that the resultant increase in clay surface polarity is accurately captured by the developed CG model. We found that a new L-type bead with reduced MARTINI LJ well-depth ($\varepsilon_{i,j}$) is required to accurately model the dispersive contribution to surface energy of phyllosilicates which have a dense covalent bonding framework. We further found that partitioning of each $\varepsilon_{i,j}$ into polar and dispersive components, and downscaling only the dispersive component provides best agreement for partitioning of TMA-MMT cleavage free energy among the polar and dispersive contributions. The mechanical properties of the clay particle were reproduced for the coarse-grained sheet by tuning the stiffness of harmonic bond and angle potentials. The stiffness values to correctly represent high rigidity of a clay particle was found to be much higher than typical MARTINIFF values. We then validated self-consistency and transferability of developed parameters for clay-polymer system by comparison of an extensive set of properties determined from the coarse-grained approach versus the atomistic approach for three polymers, viz. polyethylene (PE80), polypropylene (PP100), and polystyrene (PS60). We have calculated all properties separately in near-clay and far-clay regions, as the former is expected to provide a stringent check for suitability of developed CG parameters for modeling the clay-polymer system. Overall, we showed that developed CG parameters gave approximately 100 times speed-up in simulations and provide a good quantitative agreement for structural, thermodynamic, and dynamic properties calculated from the atomistic approach. The near-clay change in properties was indicative of weakly attractive surface-polymer interactions for all three polymers on TMA-MMT basal surface, with the exception of near-surface diffusion of PS60. The reduction in near surface polymer diffusion for PS60-MMT system was even larger than reported decrement for strongly attractive surface-polymer interactions. We relate this much stronger effect of surface on PS60 diffusion to associated structural rearrangement in near surface packing of polymers with a large side-chain group. We quantified this connection using diffusion-excess entropy scaling, where the latter captures static pair and higher order correlations between atom or bead positions. Data from both atomistic and coarse-grained simulations followed this dynamic-thermodynamic scaling. In future we will investigate applicability of developed parameters for hydrophilic polymer melts as well as polymers in aqueous solutions.

ASSOCIATED CONTENT

**Supporting Information**. The supporting information file contains additional method details on property calculation for a polymer-clay system, MARTINI-LJ well-depths for new CG beads and procedure for decomposition of CG potential energy in polar and dispersive contributions. SI_Khan2019CGFF_PCNC.pdf: Supporting information file (file type: PDF file)

AUTHOR INFORMATION

**Corresponding Author**

*goelg@chemical.iitd.ac.in

**Author Contributions**

The manuscript was written through contributions of all authors. All authors have given approval to the final version of the manuscript.

**Funding Sources**

P.K. thanks Ph.D. scholarship by MHRD, Government of India.

ACKNOWLEDGMENT

The authors thank IIT Delhi HPC facility for computational resources.

REFERENCES

(1) Yang, L.; Zhang, L.; Webster, T. J. Nanobiomaterials: State of the Art and Future Trends. *Adv. Eng. Mater.* **2011**, *13* (6), B197–B217.

(2) Grumezescu, A.; Holban, A. *Impact of Nanoscience in the Food Industry*; 2018.

(3) Yan, L.; Xie, X. Progress in Polymer Science Computational Modeling and Simulation of Nanoparticle Self-Assembly in Polymeric Systems : Structures , Properties and External Field Effects. *Prog. Polym. Sci.* **2013**, *38* (2), 369–405.

(4) Harmandaris, V. a; Daoulas, K. C.; Mavrantzas, V. G. Molecular Dynamics Simulation of a Polymer Melt / Solid Interface : Structure , Density , and Conformation of a Thin Film of Polyethylene Melt Adsorbed on Graphite. *Macromolecules* **2005**, *38* (13), 5780–5795.

(5) Giannelis, E. P. Polymer Layered Silicate Nanocomposites. *Adv. Mater.* **1996**, *8* (1), 29–35.

(6) Alexandre, M.; Dubois, P. Polymer-Layered Silicate Nanocomposites: Preparation, Properties and Uses of a New Class of Materials. *Mater. Sci. Eng. R Reports* **2000**, *28* (1), 1–63.

(7) Pavlidou, S.; Papaspyrides, C. D. A Review on Polymer-Layered Silicate Nanocomposites. *Prog. Polym. Sci.* **2008**, *33* (12), 1119–1198.

(8) Uyama, H.; Kuwabara, M.; Tsujimoto, T.; Nakano, M.; Usuki, A.; Kobayashi, S. Green Nanocomposites from Renewable Resources: Plant Oil-Clay Hybrid Materials. *Chem. Mater.* **2003**, *15* (13), 2492–2494.

(9) Pereira, S. P.; Scocchi, G.; Toth, R.; Posocco, P.; Nieto, D. R.; Pricl, S.; Fermeglia, M. Multiscale Modeling of Polymer/Clay Nanocomposites. *J. Multiscale Model.* **2011**, *03* (03), 151–176.

(10) Heinz, H. Clay Minerals for Nanocomposites and Biotechnology: Surface Modification, Dynamics and Responses to Stimuli. *Clay Miner.* **2012**, *47* (2), 205–230.

(11) Johnston, K.; Harmandaris, V. Hierarchical Simulations of Hybrid Polymer–solid Materials. *Soft Matter* **2013**, *9* (29), 6696–6710.

(12) Dunkerley, E.; Koerner, H.; Vaia, R. A.; Schmidt, D. Structure and Dynamic Mechanical Properties of Highly Oriented PS/Clay Nanolaminates over the Entire Composition Range. *Polymer (Guildf).* **2011**, *52* (4), 1163–1171.

(13) Dunkerley, E.; Schmidt, D. Effects of Composition, Orientation and Temperature on the O2 Permeability of Model Polymer/Clay Nanocomposites. *Macromolecules* **2010**, *43* (24), 10536–10544.

(14) Picard, E.; Gérard, J.-F.; Espuche, E. Reinforcement of the Gas Barrier Properties of Polyethylene and Polyamide Through the Nanocomposite Approach: Key Factors and Limitations. *Oil Gas Sci. Technol. – Rev. d'IFP Energies Nouv.* **2015**, *70* (2), 237–249.

(15) Picard, E.; Vermogen, A.; Gérard, J. F.; Espuche, E. Barrier Properties of Nylon 6-Montmorillonite Nanocomposite Membranes Prepared by Melt Blending: Influence of the Clay Content and Dispersion State. Consequences on Modelling. *J. Memb. Sci.* **2007**, *292* (1–2), 133–144.

(16) Durmuş, A.; Woo, M.; Kaşgöz, A.; Macosko, C. W.; Tsapatsis, M. Intercalated Linear Low Density Polyethylene (LLDPE)/Clay Nanocomposites Prepared with Oxidized Polyethylene as a New Type Compatibilizer: Structural, Mechanical and Barrier Properties. *Eur. Polym. J.* **2007**, *43* (9), 3737–3749.

(17) Bharadwaj, R. K. Modeling the Barrier Properties of Polymer-Layered Silicate Nanocomposites. *Macromolecules* **2001**.

(18) Eitzman, D. M.; Melkote, R. R.; Cussler, E. L. Barrier Membranes with Tipped Impermeable Flakes. *AIChE J.* **1996**.

(19) Vaia, R. A.; Liu, W.; Koerner, H. Analysis of Small-Angle Scattering of Suspensions of Organically Modified Montmorillonite: Implications to Phase Behavior of Polymer Nanocomposites. *J. Polym. Sci. Part B Polym. Phys.* **2003**.

(20) Cole, K. C.; Perrin-Sarazin, F.; Dorval-Douville, G. Infrared Spectroscopic Characterization of Polymer and Clay Platelet Orientation in Blown Films Based on Polypropylene-Clay Nanocomposite. In *Macromolecular Symposia*; 2005.

(21) Fredrickson, G. H.; Bicerano, J. Barrier Properties of Oriented Disk Composites. *J. Chem. Phys.* **1999**.

(22) Martini, S.; Kim, D. A.; Ollivon, M.; Marangoni, A. G. The Water Vapor Permeability of Polycrystalline Fat Barrier Films. *J. Agric. Food Chem.* **2006**, *54* (5), 1880–1886.

(23) Hagita, K.; Morita, H.; Doi, M.; Takano, H. Coarse-Grained Molecular Dynamics Simulation of Filled Polymer Nanocomposites under Uniaxial Elongation. *Macromolecules* **2016**, *49* (5), 1972–1983.

(24) De Virgiliis, a; Milchev, a; Rostiashvili, V. G.; Vilgis, T. a. Structure and Dynamics of a Polymer Melt at an Attractive Surface. *Eur. Phys. J. E. Soft Matter* **2012**, *35* (9), 97.

(25) Peters, B. L.; Salerno, K. M.; Agrawal, A.; Perahia, D.; Grest, G. S. Coarse-Grained

Modeling of Polyethylene Melts: Effect on Dynamics. *J. Chem. Theory Comput.* **2017**, *13* (6), 2890–2896.

(26) Wang, Q.; Keffer, D. J.; Nicholson, D. M. A Coarse-Grained Model for Polyethylene Glycol Polymer. *J. Chem. Phys.* **2011**, *135* (21), 214903.

(27) Akkermans, R. L. C.; Briels, W. J. A Structure-Based Coarse-Grained Model for Polymer Melts. *J. Chem. Phys.* **2001**, *114* (2), 1020–1031.

(28) Panizon, E.; Bochicchio, D.; Monticelli, L.; Rossi, G. MARTINI Coarse-Grained Models of Polyethylene and Polypropylene. *J. Phys. Chem. B* **2015**, *119* (25), 8209–8216.

(29) Rossi, G.; Monticelli, L.; Puisto, S. R.; Vattulainen, I.; Ala-Nissila, T. Coarse-Graining Polymers with the MARTINI Force-Field: Polystyrene as a Benchmark Case. *Soft Matter* **2011**, *7* (2), 698–708.

(30) Wang, Y.; Jiang, W. E. I.; Yan, T.; Voth, G. a. Understanding Ionic Liquids through Atomistic and Coarse-Grained Molecular Dynamics Simulations. *Acc. Chem. Res.* **2007**, *40* (11), 1193–1199.

(31) Wang, H.; Junghans, C.; Kremer, K. Comparative Atomistic and Coarse-Grained Study of Water: What Do We Lose by Coarse-Graining? *Eur. Phys. J. E* **2009**, *28* (2), 221–229.

(32) Roux, B. The Calculation of the Potential of Mean Force Using Computer Simulations. *Comput. Phys. Commun.* **1995**, *91* (1–3), 275–282.

(33) Suter, J. L.; Groen, D.; Coveney, P. V. Chemically Specific Multiscale Modeling of Clay-Polymer Nanocomposites Reveals Intercalation Dynamics, Tactoid Self-Assembly and Emergent Materials Properties. *Adv. Mater.* **2015**, *27* (6), 966–984.

(34) Sheng, N.; Boyce, M. C.; Parks, D. M.; Rutledge, G. C.; Abes, J. I.; Cohen, R. E. Multiscale Micromechanical Modeling of Polymer/Clay Nanocomposites and the Effective Clay Particle. *Polymer (Guildf).* **2004**, *45* (2), 487–506. 0

(35) Scocchi, G.; Posocco, P.; Fermeglia, M.; Pricl, S. Polymer - Clay Nanocomposites: A Multiscale Molecular Modeling Approach. *J. Phys. Chem. B* **2007**, *111* (9), 2143–2151.

(36) Liu, J.; Zhang, L.; Cao, D.; Wang, W. Static, Rheological and Mechanical Properties of Polymer Nanocomposites Studied by Computer Modeling and Simulation. *Phys. Chem. Chem. Phys.* **2009**, *11* (48), 11365–11384.

(37) Carbone, P.; Varzaneh, H. A. K.; Chen, X.; Müller-Plathe, F. Transferability of Coarse-Grained Force Fields: The Polymer Case. *J. Chem. Phys.* **2008**, *128* (6), 064904.

(38) Qian, H. J.; Carbone, P.; Xiaoyu, C.; Karimi-Varzaneh, H. A.; Liew, C. C.; Müller-Plathe, F. Temperature-Transferable Coarse-Grained Potentials for Ethylbenzene, Polystyrene, and Their Mixtures. *Macromolecules* **2008**, *41* (24), 9919–9929.

(39) Marrink, S. J.; de Vries, A. H.; Mark, A. E. Coarse Grained Model for Semiquantitative Lipid Simulations. *J. Phys. Chem. B* **2004**, *108* (2), 750–760.

(40) Marrink, S. J.; Risselada, H. J.; Yefimov, S.; Tieleman, D. P.; De Vries, A. H. The MARTINI Force Field: Coarse Grained Model for Biomolecular Simulations. *J. Phys. Chem. B* **2007**, *111* (27), 7812–7824.

(41) Buslaev, P.; Gushchin, I. Effects of Coarse Graining and Saturation of Hydrocarbon Chains on Structure and Dynamics of Simulated Lipid Molecules. *Sci. Rep.* **2017**, *7*, 11476.

(42) Lee, H.; De Vries, A. H.; Marrink, S. J.; Pastor, R. W. A Coarse-Grained Model for Polyethylene Oxide and Polyethylene Glycol: Conformation and Hydrodynamics. *J. Phys. Chem. B* **2009**, *113* (40), 13186–13194.

(43) Wallace, E. J.; Sansom, M. S. P. Carbon Nanotube/Detergent Interactions via Coarse-Grained Molecular Dynamics. *Nano Lett.* **2007**, *7* (7), 1923–1928.

(44) Gobbo, C.; Beurroies, I.; Ridder, D. De; Eelkema, R.; Marrink, S. J.; Feyter, S. De; Esch, J. H. Van; Vries, A. H. De. MARTINI Model for Physisorption of Organic Molecules on Graphite. *J. Phys. Chem. C* **2013**, *117* (30), 15623–15631.

(45) Kondela, T.; Gallov, J.; Hauß, T.; Barnoud, J.; Marrink, S. Alcohol Interactions with Lipid Bilayers. *Molecules* **2017**, *22* (12), 1–15.

(46) Heinz, H.; Koerner, H.; Anderson, K. L.; Vaia, R. A.; Farmer, B. L.; Uni, V.; May, R. V; Re, V.; Recei, M.; August, V. Force Field for Mica-Type Silicates and Dynamics of Octadecylammonium Chains Grafted to Montmorillonite. *Chem. Mater.* **2005**, *17* (5), 5658–5669.

(47) Newton, A. G.; Kwon, K. D.; Cheong, D. Edge Structure of Montmorillonite from Atomistic Simulations. *Minerals* **2016**, *6* (25), 1–15.

(48) Fetters, L. J.; Lohse, D. J.; Colby, R. H. Chain Dimensions and Entanglement Spacings. In *Physical Properties of Polymers Handbook*; 2007; pp 447–454.

(49) Lindahl, E.; Hess, B.; van der Spoel, D. GROMACS 3.0: A Package for Molecular Simulation and Trajectory Analysis. *Journal of Molecular Modeling*. 2001, pp 306–317.

(50) Van Der Spoel, D.; Lindahl, E.; Hess, B.; Groenhof, G.; Mark, A. E.; Berendsen, H. J. C.

GROMACS: Fast, Flexible, and Free. *Journal of Computational Chemistry*. 2005, pp 1701–1718.

(51) Martinez, L.; Andrade, R.; Birgin, E. G.; Martínez, J. M. PACKMOL: A Package for Building Initial Configurations for Molecular Dynamics Simulations. *J. Comput. Chem.* **2009**, *30* (13), 2157–2164.

(52) Huang, J.; Mackerell, A. D. CHARMM36 All-Atom Additive Protein Force Field: Validation Based on Comparison to NMR Data. *J. Comput. Chem.* **2013**, *34* (25), 2135–2145.

(53) Huang, J.; Rauscher, S.; Nawrocki, G.; Ran, T.; Feig, M.; De Groot, B. L.; Grubmüller, H.; MacKerell, A. D. CHARMM36m: An Improved Force Field for Folded and Intrinsically Disordered Proteins. *Nat. Methods* **2016**, *14* (1), 71–73.

(54) Heinz, H.; Lin, T. J.; Kishore Mishra, R.; Emami, F. S. Thermodynamically Consistent Force Fields for the Assembly of Inorganic, Organic, and Biological Nanostructures: The INTERFACE Force Field. *Langmuir* **2013**, *29* (6), 1754–1765.

(55) Marrink, S. J.; Risselada, H. J.; Yefimov, S.; Tieleman, D. P.; de Vries, A. H. The MARTINI Force Field: Coarse Grained Model for Biomolecular Simulations. *J. Phys. Chem. B* **2007**, *111* (27), 7812–7824.

(56) De Jong, D. H.; Baoukina, S.; Ingólfsson, H. I.; Marrink, S. J. Martini Straight: Boosting Performance Using a Shorter Cutoff and GPUs. *Comput. Phys. Commun.* **2016**, *199*, 1–7.

(57) Berendsen, H. J. C.; Postma, J. P. M.; Van Gunsteren, W. F.; Dinola, A.; Haak, J. R. Molecular Dynamics with Coupling to an External Bath. *J. Chem. Phys.* **1984**, *81* (8), 3684–3690.

(58) Parrinello, M.; Rahman, A. Polymorphic Transitions in Single Crystals: A New Molecular Dynamics Method. *J. Appl. Phys.* **1981**, *52*, 7182–7190.

(59) Lekhnitskiy, S. G. *Anisotropic Plates*; 1969.

(60) Manevitch, O. L.; Rutledge, G. C. Elastic Properties of a Single Lamella of Montmorillonite by Molecular Dynamics Simulation. *J. Phys. Chem. B* **2004**, *108* (4), 1428–1435.

(61) Heinz, H.; Vaia, R. a; Farmer, B. L. Interaction Energy and Surface Reconstruction between Sheets of Layered Silicates. *J. Chem. Phys.* **2006**, *124* (22), 224713.

(62) Van Oss, C. J.; Chaudhury, M. K.; Good, R. J. Interfacial Lifshitz-van Der Waals and Polar Interactions in Macroscopic Systems. *Chem. Rev.* **1988**, *88* (6), 927–941.

(63) Giese, R. F.; Van Oss, C. J. *Colloid And Surface Properties Of Clays And Related Minerals*; 2002.

(64) Perrin, E.; Schoen, M.; Coudert, F. X.; Boutin, A. Structure and Dynamics of Solvated Polymers near a Silica Surface: On the Different Roles Played by Solvent. *J. Phys. Chem. B* **2018**, *122* (16), 4573−4582.

(65) Heinz, H.; Suter, U. W. Atomic Charges for Classical Simulations of Polar Systems. *J. Phys. Chem. B* **2004**, *108* (47), 18341–18352.

(66) Doi, M.; Edwards, S. F. *The Theory of Polymer Dynamics*; 1986.

(67) Borzsák, I.; Baranyai, A. On the Convergence of Green's Entropy Expansion. *Chem. Phys.* **1992**, *165* (2–3), 227–230.

(68) Laird, B. B.; Haymet, A. D. J. Calculation of the Entropy of Binary Hard Sphere Mixtures from Pair Correlation Functions. *J. Chem. Phys.* **1992**, *97* (3), 2153–2155.

(69) Schlitter, J. Estimation of Absolute and Relative Entropies of Macromolecules Using the Covariance Matrix. *Chem. Phys. Lett.* **1993**, *215* (6), 617–621.

(70) Rosenfeld, Y. Relation between the Transport Coefficients and the Internal Entropy of Simple Systems. *Phys. Rev. A* **1977**, *15* (6), 2545–2549.

(71) Rosenfeld, Y. A Quasi-Universal Scaling Law for Atomic Transport in Simple Fluids A Quasi-Universal Scaling Law for Atomic Transport in Simple Fluids. *J.Phys.F:Metal. Phys.* **1999**, *11* (28), 5415–5427.

(72) Pönitzsch, A.; Nofz, M.; Wondraczek, L.; Deubener, J. Bulk Elastic Properties, Hardness and Fatigue of Calcium Aluminosilicate Glasses in the Intermediate-Silica Range. *J. Non. Cryst. Solids* **2016**, *434*, 1–12.

(73) Sinclair, W.; Ringwood, A. E. Single Crystal Analysis of the Structure of Stishovite. *Nature* **1978**, *272* (5655), 714–715.

(74) Suter, J. L.; Coveney, P. V.; Greenwell, H. C.; Thyveetil, M. A. Large-Scale Molecular Dynamics Study of Montmorillonite Clay: Emergence of Undulatory Fluctuations and Determination of Material Properties. *J. Phys. Chem. C* **2007**, *111* (23), 8248–8259.

(75) Stark, A. C.; Andrews, C. T.; Elcock, A. H. Toward Optimized Potential Functions for Protein-Protein Interactions in Aqueous Solutions: Osmotic Second Virial Coefficient Calculations Using the MARTINI Coarse-Grained Force Field. *J. Chem. Theory Comput.* **2013**, *9* (9), 4176–4185.

(76) Schmalhorst, P. S.; Deluweit, F.; Scherrers, R.; Heisenberg, C. P.; Sikora, M. Overcoming the Limitations of the MARTINI Force Field in Simulations of Polysaccharides. *J. Chem. Theory Comput.* **2017**, *13* (10), 5039–5053.

(77) Ricci, C. G.; Li, B.; Cheng, L. T.; Dzubiella, J.; McCammon, J. A. "Martinizing" the Variational Implicit Solvent Method (VISM): Solvation Free Energy for Coarse-Grained Proteins. *J. Phys. Chem. B* **2017**, *121* (27), 6538–6548.

(78) Dee, G. T.; Ougizawa, T.; Walsh, D. J. The Pressure-Volume-Temperature Properties of Polyethylene, Poly(Dimethyl Siloxane), Poly(Ethylene Glycol) and Poly(Propylene Glycol) as a Function of Molecular Weight. *Polymer (Guildf).* **1992**, *33* (16), 3462–3469.

(79) Sato, Y.; Yamasaki, Y.; Takishima, S.; Masuoka, H. Precise Measurement of the PVT of Polypropylene and Polycarbonate up to 330°C and 200 MPa. *J. Appl. Polym. Sci.* **1997**, *66* (1), 141–150.

(80) Utracki, L. A.; Kamal, M. R. *Clay-Containing Polymeric Nanocomposites*; 2002; Vol. 27.

(81) Suter, J. L.; Coveney, P. V. Computer Simulation Study of the Materials Properties of Intercalated and Exfoliated Poly(Ethylene)Glycol Clay Nanocomposites. *Soft Matter* **2009**, *5* (11), 2239.

(82) Ndoro, T. V. M.; Voyiatzis, E.; Ghanbari, A.; Theodorou, D. N.; Böhm, M. C.; Müller-Plathe, F. Interface of Grafted and Ungrafted Silica Nanoparticles with a Polystyrene Matrix: Atomistic Molecular Dynamics Simulations. *Macromolecules* **2011**, *44* (7), 2316–2327.

(83) Smit, B.; Karabomi, S.; Ilja Siepmann, J. Computer Simulations of Vapor-Liquid Phase Equilibria of n-Alkanes. *J. Chem. Phys.* **1995**, *102* (5), 2126–2140.

(84) Ballard, D. G. H.; Cheshire, P.; Longman, G. W.; Schelten, J. Small-Angle Neutron Scattering Studies of Isotropic Polypropylene. *Polymer (Guildf).* **1978**, *19* (4), 379–385.

(85) Konishi, T.; Yoshizaki, T.; Saito, T.; Einaga, Y.; Yamakawa, H. Mean-Square Radius of Gyration of Oligo- and Polystyrenes in Dilute Solutions. *Macromolecules* **1990**, *23* (1), 290–297.

(86) Kraus, J.; Muller-Buschbaum, P.; Kuhlmann, T.; Schubert, D. W.; Stamm, M. Confinement Effects on the Chain Conformation in Thin Polymer Films. *Eur. Lett* **2000**, *49* (2), 210–216.

(87) Sikorski, A. Properties of Confined Polymer Melts. *ACTA Phys. Pol. A* **2005**, *107* (3), 443–450.

(88) Cho, S.; Jeong, S.; Kim, J. M.; Baig, C. Molecular Dynamics for Linear Polymer Melts in Bulk and Confined Systems under Shear Flow. *Sci. Rep.* **2017**, *7* (1), 1–9.

(89) Goel, G.; Krekelberg, W. P.; Errington, J. R.; Truskett, T. M. Tuning Density Profiles and Mobility of Inhomogeneous Fluids. *Phys. Rev. Lett.* **2008**, *100* (10).

(90) Ying, R.; Jian, G.; Wei, G.; JingHai, L.; GuoHua, H. Molecular Dynamics Simulation of a Single Polymer in Hydrophilic Nano-Slits. *Chinese Sci. Bull.* **2008**, *53* (17), 2599–2606. 9.

(91) Nowicki, W.; Nowicka, G.; Narkiewicz-Michałek, J. Influence of Confinement on Conformational Entropy of a Polymer Chain and Structure of Polymer-Nanoparticles Complexes. *Polymer (Guildf)*. **2009**, *50* (9), 2161–2171.

(92) Bachus R. and Kimmich R. Molecular Weight and Temperature Dependence of Self-Diffusion Coefficients in Polyethylene and Polystyrene Melts Investigated Using a Modified n.m.r, Fieldgradient Technique. *Polymer (Guildf)*. **1983**, *24* (8), 964–970.

(93) Desai, T.; Keblinski, P.; Kumar, S. K. Molecular Dynamics Simulations of Polymer Transport in Nanocomposites. *J. Chem. Phys.* **2005**, *122* (13), 1–8.